\documentclass[showpacs,aps,prd]{revtex4}
\input epsf

\begin{document}

\title{An open charm tetraquark candidate: note on $X_{0}(2900)$}
\author{Jian-Rong Zhang}
\affiliation{Department of Physics, College of Liberal Arts and Sciences, National University of Defense Technology,
Changsha 410073, Hunan, People's Republic of China}


\begin{abstract}
Motivated by the LHCb's very recent observation of exotic $X_{0}(2900)$ in the
$B^{+}\rightarrow D^{+}D^{-}K^{+}$ process,
for which could be a good open charm $ud\bar{c}\bar{s}$ tetraquark candidate,
we endeavor
to investigate its possibility by means of QCD sum rules.
In technique, four configurations of interpolating currents with $J^{P}=0^{+}$ are
studied for the $ud\bar{c}\bar{s}$ tetraquark state.
In the end, mass values are calculated to be $2.76^{+0.16}_{-0.23}~\mbox{GeV}$
for the axial vector diquark-axial vector antidiquark configuration and
$2.75^{+0.15}_{-0.24}~\mbox{GeV}$
for the scalar diquark-scalar antidiquark configuration,
both of which are consistent with the experimental data $2.866\pm0.007\pm0.002~\mbox{GeV}$ of $X_{0}(2900)$
in view of the uncertainty.
These results support that $X_{0}(2900)$ could be a $0^{+}$ tetraquark state with open charm flavor.
\end{abstract}
\maketitle

\section{Introduction}\label{sec1}

In the past decades, the so-called $X$, $Y$, and $Z$ new hadrons
have attracted wide attentions and some of them were assigned to be possible exotic states
(for recent reviews e.g. see \cite{Th-rev,Th-rev1,Th-rev2} and references therein).
For example, some prediction was made
on a $0^{+}$ bound
state with a pole mass of $2848~\mbox{MeV}$
via coupled-channel unitarity \cite{Oset}.
Very recently, the LHCb Collaboration reported
the first amplitude analysis of the $B^{+}\rightarrow D^{+}D^{-}K^{+}$ decay
and included two new exotic structures in
the $D^{-}K^{+}$ channel
with an overwhelming significance \cite{X2900}.
Particularly for the spin-$0$ resonance $X_{0}(2900)$,
its mass and width were measured to be
 $2.866\pm0.007\pm0.002~\mbox{GeV}$ and
$57\pm12\pm4~\mbox{MeV}$, respectively.
Taking notice of its decay final states being $D^{-}K^{+}$,
$X_{0}(2900)$ was proposed to be a nice open charm
tetraquark candidate \cite{Th}.

Activated by the LHCb's new experimental result on $X_{0}(2900)$,
we attempt to study its possibility to be an open charm $ud\bar{c}\bar{s}$ tetraquark state.
To research into a genuine hadron, one has to face the very complicated nonperturbative QCD problem.
As one trustable approach for evaluating
nonperturbative effects, the QCD sum rule \cite{svzsum} is
firmly established on QCD basic theory and
has been successfully applied to numerous hadronic systems \cite{overview1,overview2,overview3,reinders,overview4}.
For instance, the charm-strange
$D_{s0}^{*}(2317)$ was explored in a tetraquark picture
with QCD sum rules \cite{tetra1,tetra2,tetra3-heavy-limit,tetra3-Narison,tetra3,tetra4,tetra5}.
In this work, to uncover the internal structure of $X_{0}(2900)$,
we devote to investigating that whether it could be an open charm $ud\bar{c}\bar{s}$ tetraquark state
by QCD sum rules.

The rest of the paper is organized as follows. In Sec. \ref{sec2},
$X_{0}(2900)$ is
studied as a tetraquark state via QCD sum
rules, followed by numerical analysis and discussions in Sec. \ref{sec3}. The last part gives a concise
summary.
\section{QCD sum rule study of $X_{0}(2900)$ as a $0^{+}$ $ud\bar{c}\bar{s}$ tetraquark state}\label{sec2}
Complying with the usual treatment of QCD sum rules, a tetraquark state can be represented by
an interpolating current with the
diquark-antidiquark configuration (e.g. see the review \cite{overview4}
and references therein).
Particularly for the present $0^{+}$ $ud\bar{c}\bar{s}$ tetraquark state,
one can construct its different configuration currents with $0^{+}$ composed of an $ud$-diquark
and a $\bar{c}\bar{s}$-antidiquark, taking into account that
$u_{a}^{T}C\gamma_{5}d_{b}$ as
a $0^{+}$ scalar diquark,
$u_{a}^{T}Cd_{b}$ as a $0^{-}$ pseudoscalar diquark,
$u_{a}^{T}C\gamma_{\mu}d_{b}$ as
a $1^{+}$ axial vector diquark, $u_{a}^{T}C\gamma_{5}\gamma_{\mu}d_{b}$
as a $1^{-}$ vector diquark, and likewise for the $\bar{c}\bar{s}$-antidiquark.
In this manner,
following forms of currents are presented for the $0^{+}$ $ud\bar{c}\bar{s}$ tetraquark state, with
\begin{eqnarray}
j_{(1)}=\epsilon_{abg}\epsilon_{a'b'g}(u_{a}^{T}C\gamma_{5}d_{b})(\bar{c}_{a'}\gamma_{5}C\bar{s}_{b'}^{T})
\end{eqnarray}
for the scalar diquark-scalar antidiquark configuration,
\begin{eqnarray}
j_{(2)}&=&\epsilon_{abg}\epsilon_{a'b'g}(u_{a}^{T}Cd_{b})(\bar{c}_{a'}C\bar{s}_{b'}^{T})
\end{eqnarray}
for the pseudoscalar diquark-pseudoscalar antidiquark configuration,
\begin{eqnarray}
j_{(3)}&=&\epsilon_{abg}\epsilon_{a'b'g}(u_{a}^{T}C\gamma_{\mu}d_{b})(\bar{c}_{a'}\gamma^{\mu}C\bar{s}_{b'}^{T})
\end{eqnarray}
for the axial vector diquark-axial vector antidiquark configuration, and
\begin{eqnarray}
j_{(4)}&=&\epsilon_{abg}\epsilon_{a'b'g}(u_{a}^{T}C\gamma_{5}\gamma_{\mu}d_{b})(\bar{c}_{a'}\gamma^{\mu}\gamma_{5}C\bar{s}_{b'}^{T})
\end{eqnarray}
for the vector diquark-vector antidiquark configuration.
Here $T$ indicates matrix transposition, $C$ is the charge conjugation matrix, and
the subscripts $a$,
$b$, $g$, $a'$, and $b'$ are color indices.

Besides, one can construct some other current, such as
\begin{eqnarray}
j_{(5)}&=&\epsilon_{abg}\epsilon_{a'b'g}(u_{a}^{T}C\sigma_{\mu\nu}d_{b})(\bar{c}_{a'}\sigma^{\mu\nu}C\bar{s}_{b'}^{T}).
\end{eqnarray}
Meanwhile, one could consider that the
corresponding diquark and antidiquark are
higher excitation and they are comparatively
difficult to be stably formed. Then the concrete calculations for this current are not involved here,
for which could be taken into account in some further work.

On the one hand, the two-point correlator
\begin{eqnarray}
\Pi_{i}(q^{2})=i\int
d^{4}x\mbox{e}^{iq.x}\langle0|T[j_{(i)}(x)j_{(i)}^{\dag}(0)]|0\rangle, ~(i=1,~2,~3,~\mbox{or}~4)
\end{eqnarray}
can be phenomenologically expressed as
\begin{eqnarray}\label{ph}
\Pi_{i}(q^{2})=\frac{\lambda_{H}^{2}}{M_{H}^{2}-q^{2}}+\frac{1}{\pi}\int_{s_{0}}
^{\infty}\frac{\mbox{Im}\big[\Pi_{i}^{\mbox{phen}}(s)\big]}{s-q^{2}}ds,
\end{eqnarray}
where $s_0$ is the continuum threshold,
$M_{H}$ is the hadron's mass, and $\lambda_{H}$ denotes
the hadronic coupling constant  $\langle0|j|H\rangle=\lambda_{H}$.

On the other hand, $\Pi_{i}(q^{2})$ can be theoretically formalized as
\begin{eqnarray}\label{ope}
\Pi_{i}(q^{2})=\int_{(m_{c}+m_{s})^{2}}^{\infty}\frac{\rho_{i}}{s-q^{2}}ds+\Pi_{i}^{\mbox{cond}}(q^{2}),
\end{eqnarray}
in which $m_{c}$ is the charm mass, $m_{s}$ is the strange mass,
and the spectral density $\rho_{i}=\frac{1}{\pi}\mbox{Im}\big[\Pi_{i}(s)\big]$.

Matching the two equations (\ref{ph}) and (\ref{ope}), assuming quark-hadron duality, and
making a Borel transform $\hat{B}$, the sum rule can be written as
\begin{eqnarray}\label{sumrule1}
\lambda_{H}^{2}e^{-M_{H}^{2}/M^{2}}&=&\int_{(m_{c}+m_{s})^{2}}^{s_{0}}\rho_{i} e^{-s/M^{2}}ds+\hat{B}\Pi_{i}^{\mbox{cond}},
\end{eqnarray}
with the Borel parameter $M^2$.

Taking the derivative of the sum rule (\ref{sumrule1}) with $-\frac{1}{M^{2}}$
and then dividing the result by (\ref{sumrule1}) itself, one can get the hadronic
mass
\begin{eqnarray}\label{sum rule 1}
M_{H}=\sqrt{\bigg[\int_{(m_{c}+m_{s})^{2}}^{s_{0}}\rho_{i} s
e^{-s/M^{2}}ds+\frac{d\big(\hat{B}\Pi_{i}^{\mbox{cond}}\big)}{d(-\frac{1}{M^{2}})}\bigg]/
\bigg[\int_{(m_{c}+m_{s})^{2}}^{s_{0}}\rho_{i} e^{-s/M^{2}}ds+\hat{B}\Pi_{i}^{\mbox{cond}}\bigg]},
\end{eqnarray}
for which the spectral density $\rho_{i}$ and the term $\hat{B}\Pi_{i}^{\mbox{cond}}$
can be derived making use of the similar techniques as Refs. \cite{tetra1,tetra2,overview4,Zhang},
and for concision their expressions are wholly enclosed in the Appendix A.

One could note that
there is not
the $\langle\bar{q}q\rangle$ or $\langle g\bar{q}\sigma\cdot G q\rangle$
condensate in this work,
which is mainly resulted from two aspects of reasons.
On the one hand, both light $u$ and $d$ quark masses are so small comparing with
the heavy charm mass that light quark masses have been safely neglected,
and thus there does not appear the $\langle\bar{q}q\rangle$ or $\langle g\bar{q}\sigma\cdot G q\rangle$ term
proportional to $m_{u}$ or $m_{d}$.
On the other hand, the $\langle\bar{q}q\rangle$ or $\langle g\bar{q}\sigma\cdot G q\rangle$ term
without $m_{u}$ or $m_{d}$ vanishes owing to that
its corresponding matrix trace happens to be zero.
In this way, the spectral density $\rho$ does not contain
the $\langle\bar{q}q\rangle$ or $\langle g\bar{q}\sigma\cdot G q\rangle$ condensate.

\section{numerical analysis and discussions}\label{sec3}
To extract the mass $M_{H}$,
one can carry out the numerical analysis of sum rule (\ref{sum rule 1}),
with the aid of input parameters
$\langle\bar{q}q\rangle=-(0.24\pm0.01)^{3}~\mbox{GeV}^{3}$,
$\langle g\bar{q}\sigma\cdot G q\rangle=m_{0}^{2}~\langle\bar{q}q\rangle$,
$\langle\bar{s}s\rangle=m_{0}^{2}~\langle\bar{q}q\rangle$,
$\langle g\bar{s}\sigma\cdot G s\rangle=m_{0}^{2}~\langle\bar{s}s\rangle$,
$m_{0}^{2}=0.8\pm0.1~\mbox{GeV}^{2}$, $\langle
g^{2}G^{2}\rangle=0.88\pm0.25~\mbox{GeV}^{4}$, and $\langle
g^{3}G^{3}\rangle=0.58\pm0.18~\mbox{GeV}^{6}$ \cite{svzsum,overview2}.
Besides, quark masses are taken as $m_{c}=1.27\pm0.02~\mbox{GeV}$
and $m_{s}=93^{+11}_{-5}~\mbox{MeV}$ \cite{PDG}, respectively.
Keeping to the procedure of sum rule analysis,
both the OPE convergence and pole dominance should be inspected
to find suitable work windows for the threshold $\sqrt{s_{0}}$ and the Borel
parameter $M^{2}$.

Taking the axial vector diquark-axial vector antidiquark case as an example, its
various relative OPE contributions
are compared in FIG. 1,
which displays that
there are three main condensate contributions, i.e. the two-quark condensate $\langle\bar{s}s\rangle$,
the mixed condensate $\langle\bar{s}g\sigma\cdot G s\rangle$, and the four-quark condensate $\langle\bar{q}q\rangle^{2}$.
Comparatively, one could note that
the $\langle\bar{q}q\rangle^{2}$ contribution
is bigger than the lower dimension condensate like $\langle\bar{s}s\rangle$ or $\langle\bar{s}g\sigma\cdot G s\rangle$.
Frankly speaking, this is a common problem existing in some
multiquark QCD sum rule studies but not newly arisen,
for which has already been discussed in some other
works \cite{P-Matheus,P-Chen,P-Wang,P-Zhang}. Namely,
some individual high dimension condensate (e.g. the  $\langle\bar{q}q\rangle^{2}$ here) plays an
important role on the OPE side, which causes that it
is not easy to satisfy the traditional condition for conventional
hadrons that low dimension condensate should be
bigger than high dimension one in the OPE.
Acceptably, these main condensate contributions
can fortunately counteract each other
to some extent.
One might note that the $\langle\bar{q}q\rangle^{2}\langle\bar{s}s\rangle$
condensate is not small at $M^{2}=1~\mbox{GeV}^{2}$, however,
it descends rapidly with the increase of $M^{2}$ and becomes very small
while taking $M^{2}\geq2.0~\mbox{GeV}^{2}$.
All these factors bring that the lowest dimension perturbative part
can play an important role on the total OPE
when $M^{2}\geq2.0~\mbox{GeV}^{2}$
and the corresponding OPE convergence is still under control in the work windows.

In phenomenology, FIG. 2 shows
the comparison between pole contribution and
continuum contribution of sum rule (\ref{sumrule1}) for
$\sqrt{s_{0}}=3.4~\mbox{GeV}$, which manifests that the relative pole
contribution is close to $50\%$ at $M^{2}=2.2~\mbox{GeV}^{2}$
and decreases with $M^{2}$. Thereby, the upper bound of $M^{2}$ is
chosen as $2.2~\mbox{GeV}^{2}$ for
$\sqrt{s_{0}}=3.4~\mbox{GeV}$.
Similarly, the upper values of $M^{2}$ can also
be achieved for $\sqrt{s_0}=3.3~\mbox{GeV}$ and $\sqrt{s_0}=3.5~\mbox{GeV}$.
Accordingly, work windows for the axial vector diquark-axial vector antidiquark case are fixed as
$M^{2}=2.0\sim2.1~\mbox{GeV}^{2}$ for $\sqrt{s_0}=3.3~\mbox{GeV}$,
$M^{2}=2.0\sim2.2~\mbox{GeV}^{2}$ for
$\sqrt{s_0}=3.4~\mbox{GeV}$, and
$M^{2}=2.0\sim2.3~\mbox{GeV}^{2}$ for $\sqrt{s_0}=3.5~\mbox{GeV}$, respectively.
In FIG. 3, the mass $M_{H}$
as a function of $M^2$ from sum rule (\ref{sum rule 1}) is
shown for the axial vector diquark-axial vector antidiquark case.
Within the chosen work windows, it may seem not very flat for the Borel curves.
As one knows, in the choice of work windows,
the flatness of Borel curves is an important factor under consideration.
Meanwhile, it should not be the only judgement.
In fact, the Borel curves can look much flatter if
one naively chooses Borel windows with some larger $M^{2}$ here.
However, the hypothesis of pole dominance
in the phenomenological side of QCD sum rules
would be severely broken if overly paying attention to the flatness of Borel curves.
In practice, the procedure of finding work windows
has actually been developed from
the traditional way of mainly observing that wether
there is some flat Borel plateau to the present-day way
of choosing suitable work windows fulfilling both the OPE convergence and pole dominance,
to ensure that two sides of QCD sum rules have a good overlap and
information on the hadronic resonance can be reliably extracted.
Moreover, the variation of mass with Borel parameter $M^{2}$
in the Borel curve
can be numerically embodied by the uncertainty of final result.
After considering the uncertainty from the variation of QCD parameters,
one gains the final mass $2.76^{+0.16}_{-0.23}~\mbox{GeV}$
for the axial vector diquark-axial vector antidiquark case.

\begin{figure}
\centerline{\epsfysize=7.18truecm\epsfbox{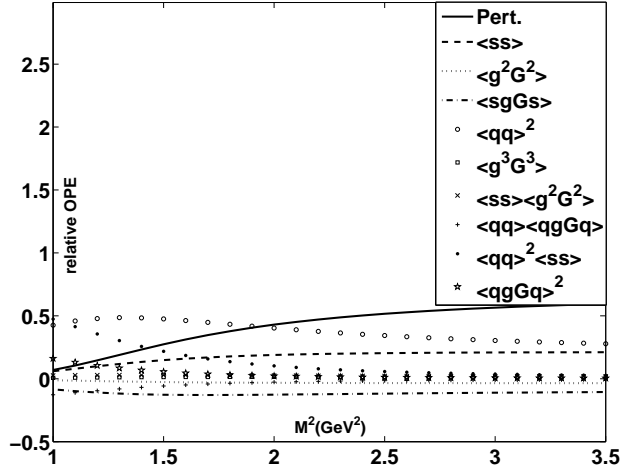}}
\caption{The relative contributions of various condensates
as a function of $M^2$ in sum rule
(\ref{sumrule1}) for $\sqrt{s_{0}}=3.4~\mbox{GeV}$ for the axial vector diquark-axial vector antidiquark case.}
\end{figure}

\begin{figure}
\centerline{\epsfysize=7.18truecm\epsfbox{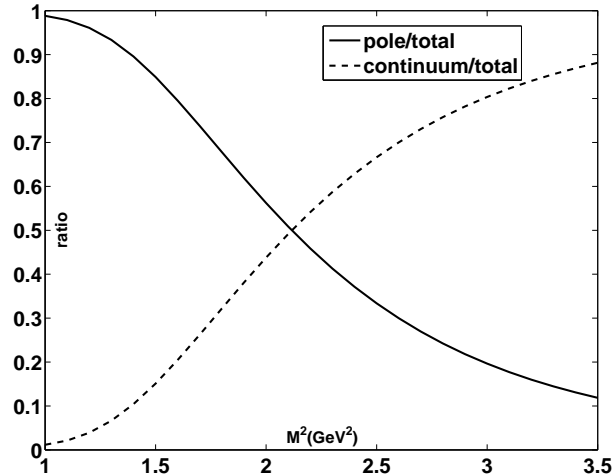}}
\caption{The phenomenological contribution as a function of $M^2$ in sum rule
(\ref{sumrule1}) for $\sqrt{s_{0}}=3.4~\mbox{GeV}$ for the axial vector diquark-axial vector antidiquark case.
The solid line is the relative pole contribution
and the dashed line is the relative continuum
contribution.}
\end{figure}

\begin{figure}
\centerline{\epsfysize=7.18truecm
\epsfbox{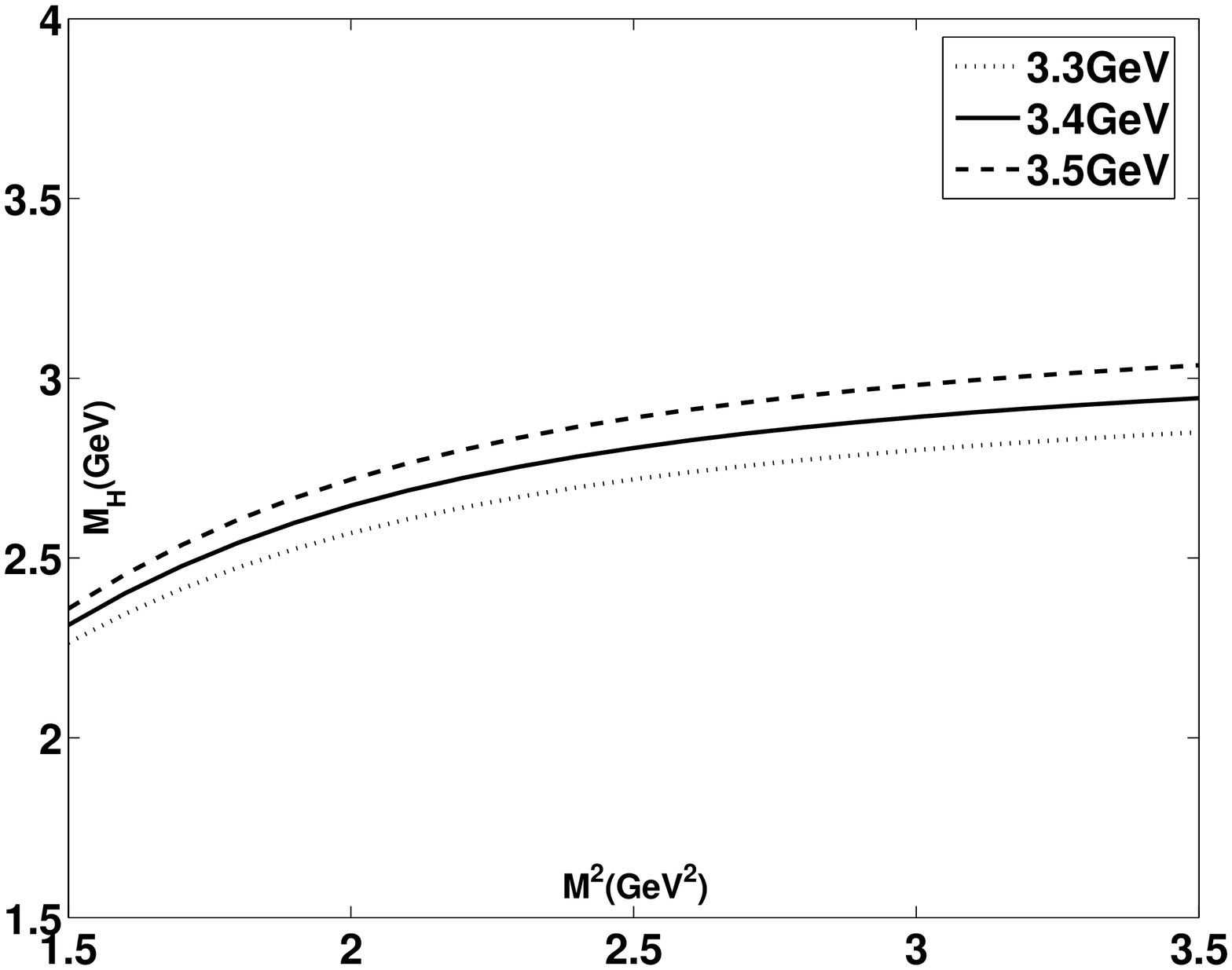}}\caption{
The mass of $0^{+}$ $ud\bar{c}\bar{s}$ tetraquark state with the axial vector diquark-axial vector antidiquark configuration as
a function of $M^2$ from sum rule (\ref{sum rule 1}). The
ranges of $M^{2}$ are taken as $2.0\sim2.1~\mbox{GeV}^{2}$ for
$\sqrt{s_0}=3.3~\mbox{GeV}$, $2.0\sim2.2~\mbox{GeV}^{2}$ for
$\sqrt{s_0}=3.4~\mbox{GeV}$, and $2.0\sim2.3~\mbox{GeV}^{2}$ for
$\sqrt{s_0}=3.5~\mbox{GeV}$, respectively.}
\end{figure}

For the scalar diquark-scalar antidiquark case, through the similar procedure, its
Borel windows are determined to be
$2.0\sim2.2~\mbox{GeV}^{2}$ for $\sqrt{s_0}=3.3~\mbox{GeV}$,
$2.0\sim2.3~\mbox{GeV}^{2}$ for
$\sqrt{s_0}=3.4~\mbox{GeV}$, and
$2.0\sim2.4~\mbox{GeV}^{2}$ for $\sqrt{s_0}=3.5~\mbox{GeV}$, respectively.
Furthermore, its mass $M_{H}$ dependence on
$M^2$ from sum rule (\ref{sum rule 1}) is
shown in FIG. 4.
Including the uncertainty due to QCD parameters,
the mass for the scalar diquark-scalar antidiquark configuration is calculated to be
$2.75^{+0.15}_{-0.24}~\mbox{GeV}$ at last.

\begin{figure}
\centerline{\epsfysize=7.18truecm
\epsfbox{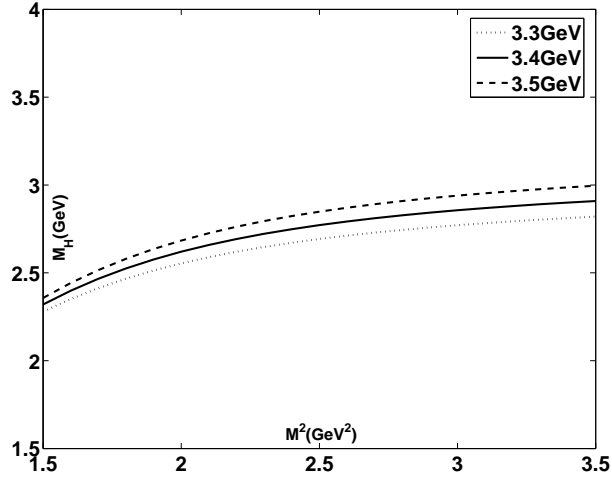}}\caption{
The mass of $0^{+}$ $ud\bar{c}\bar{s}$ tetraquark state with the scalar diquark-scalar antidiquark configuration as
a function of $M^2$ from sum rule (\ref{sum rule 1}). The
ranges of $M^{2}$ are taken as $2.0\sim2.2~\mbox{GeV}^{2}$ for
$\sqrt{s_0}=3.3~\mbox{GeV}$, $2.0\sim2.3~\mbox{GeV}^{2}$ for
$\sqrt{s_0}=3.4~\mbox{GeV}$, and $2.0\sim2.4~\mbox{GeV}^{2}$ for
$\sqrt{s_0}=3.5~\mbox{GeV}$, respectively.}
\end{figure}

After similar analysis, it is noted that the OPE convergence for
the pseudoscalar diquark-pseudoscalar antidiquark and
the vector diquark-vector antidiquark cases is so unsatisfactory
that one cannot find any appropriate work windows for them
and it is not advisable to continue extracting their mass results.
Anyway,
the final results for the axial vector diquark-axial vector antidiquark and the scalar diquark-scalar antidiquark cases
both agree with the experimental data of $X_{0}(2900)$
viewing the uncertainty of final results, which supports that $X_{0}(2900)$ could be
a $0^{+}$ $ud\bar{c}\bar{s}$ tetraquark state with the axial vector diquark-axial vector antidiquark or the scalar diquark-scalar antidiquark
configuration.

\section{Summary}\label{sec4}
Stimulated by the new observation of exotic
$X_{0}(2900)$, we explore
the possibility of
$X_{0}(2900)$ as an open charm $ud\bar{c}\bar{s}$ tetraquark state with $J^{P}=0^{+}$
in the framework of QCD sum rules.
Finally, the mass values are computed to be $2.76^{+0.16}_{-0.23}~\mbox{GeV}$
for the axial vector diquark-axial vector antidiquark configuration, and
$2.75^{+0.15}_{-0.24}~\mbox{GeV}$ for the scalar diquark-scalar antidiquark configuration, respectively.
Considering the uncertainty of these
results, they are both in agreement with the experimental data of $X_{0}(2900)$.
It supports that $X_{0}(2900)$ could be interpreted as
a $0^{+}$ $ud\bar{c}\bar{s}$ tetraquark state, whose configuration
could be either the axial vector diquark-axial vector antidiquark or the scalar diquark-scalar antidiquark.
In future, it is expected that further experimental observations and
theoretical efforts
could disclose more information on the nature of $X_{0}(2900)$.

\begin{acknowledgments}
This work was supported by the National
Natural Science Foundation of China under Contract
Nos. 11475258 and 11675263, and by the project for excellent youth talents in
NUDT.
\end{acknowledgments}


\appendix
\section{}
The spectral density $\rho_{i}=\rho_{i}^{\mbox{pert}}+\rho_{i}^{\langle\bar{s}s\rangle}+\rho_{i}^{\langle
g^{2}G^{2}\rangle}+\rho_{i}^{\langle
g\bar{s}\sigma\cdot G s\rangle}+\rho_{i}^{\langle\bar{q}q\rangle^{2}}+\rho_{i}^{\langle
g^{3}G^{3}\rangle}+\rho_{i}^{\langle\bar{s}s\rangle\langle
g^{2}G^{2}\rangle}$ and the term $\hat{B}\Pi_{i}^{\mbox{cond}}$ are collected below, with
\begin{eqnarray}
\rho_{1}^{\mbox{pert}}&=&\frac{1}{3\cdot2^{10}\pi^{6}}\int_{\Lambda}^{1}d\alpha\bigg(\frac{1-\alpha}{\alpha}\bigg)^{3}(\alpha s-m_{c}^{2}+4m_{s}m_{c})(\alpha s-m_{c}^{2})^{3},\\
\rho_{1}^{\langle\bar{s}s\rangle}&=&-\frac{\langle\bar{s}s\rangle}{2^{6}\pi^{4}}\int_{\Lambda}^{1}d\alpha\frac{1-\alpha}{\alpha^{2}}[(1-\alpha)m_{c}-\alpha m_{s}](\alpha s-m_{c}^{2})^{2},\\
\rho_{1}^{\langle g^{2}G^{2}\rangle}&=&-\frac{m_{c}\langle
g^{2}G^{2}\rangle}{3^{2}\cdot2^{10}\pi^{6}}\int_{\Lambda}^{1}d\alpha\bigg(\frac{1-\alpha}{\alpha}\bigg)^{3}[(m_{c}-3m_{s})(\alpha s-m_{c}^{2})+m_{s}m_{c}^{2}],\\
\rho_{1}^{\langle g\bar{s}\sigma\cdot G s\rangle}&=&\frac{\langle
g\bar{s}\sigma\cdot G
s\rangle}{3\cdot2^{6}\pi^{4}}\int_{\Lambda}^{1}
\frac{d\alpha}{\alpha}[3(1-\alpha)m_{c}-\alpha m_{s}]
(\alpha s-m_{c}^{2}),\\
\rho_{1}^{\langle\bar{q}q\rangle^{2}}&=&\frac{\langle\bar{q}q\rangle^{2}}{3\cdot2^{2}\pi^{2}}\int_{\Lambda}^{1}d\alpha[(\alpha s-m_{c}^{2})+m_{s}m_{c}],\\
\rho_{1}^{\langle g^{3}G^{3}\rangle}&=&-\frac{\langle
g^{3}G^{3}\rangle}{3^{2}\cdot2^{12}\pi^{6}}\int_{\Lambda}^{1}d\alpha\bigg(\frac{1-\alpha}{\alpha}\bigg)^{3}(\alpha s-3m_{c}^{2}+6m_{s}m_{c}),\\
\rho_{1}^{\langle\bar{s}s\rangle\langle
g^{2}G^{2}\rangle}&=&-\frac{m_{c}\langle\bar{s}s\rangle\langle
g^{2}G^{2}\rangle}{3^{2}\cdot2^{8}\pi^{4}}\int_{\Lambda}^{1}d\alpha\bigg[1+3\bigg(\frac{1-\alpha}{\alpha}\bigg)^{2}\bigg],\\
\hat{B}\Pi_{1}^{\mbox{cond}}&=&-\frac{m_{s}m_{Q}\langle\bar{q}q\rangle\langle g\bar{q}\sigma\cdot G q\rangle}{3\cdot2^{3}\pi^{2}}e^{-m_{Q}^{2}/M^{2}}\nonumber\\
&-&\frac{\langle\bar{q}q\rangle^{2}\langle\bar{s}s\rangle}{2\cdot3^{2}}\bigg(2m_{Q}-m_{s}-\frac{m_{s}m_{Q}^{2}}{M^{2}}\bigg)e^{-m_{Q}^{2}/M^{2}}\nonumber\\
&+&\frac{\langle g\bar{q}\sigma\cdot G q\rangle^{2}}{3\cdot2^{6}\pi^{2}}\bigg[1+\frac{m_{Q}^{2}}{M^{2}}+\frac{m_{s}m_{Q}^{3}}{(M^{2})^{2}}\bigg]e^{-m_{Q}^{2}/M^{2}},
\end{eqnarray}
for the scalar diquark-scalar antidiquark case,
\begin{eqnarray}
\rho_{2}^{\mbox{pert}}&=&\frac{1}{3\cdot2^{10}\pi^{6}}\int_{\Lambda}^{1}d\alpha\bigg(\frac{1-\alpha}{\alpha}\bigg)^{3}(\alpha s-m_{c}^{2}-4m_{s}m_{c})(\alpha s-m_{c}^{2})^{3},\\
\rho_{2}^{\langle\bar{s}s\rangle}&=&\frac{\langle\bar{s}s\rangle}{2^{6}\pi^{4}}\int_{\Lambda}^{1}d\alpha\frac{1-\alpha}{\alpha^{2}}[(1-\alpha)m_{c}+\alpha m_{s}](\alpha s-m_{c}^{2})^{2},\\
\rho_{2}^{\langle g^{2}G^{2}\rangle}&=&-\frac{m_{c}\langle
g^{2}G^{2}\rangle}{3^{2}\cdot2^{10}\pi^{6}}\int_{\Lambda}^{1}d\alpha\bigg(\frac{1-\alpha}{\alpha}\bigg)^{3}[(m_{c}+3m_{s})(\alpha s-m_{c}^{2})-m_{s}m_{c}^{2}],\\
\rho_{2}^{\langle g\bar{s}\sigma\cdot G s\rangle}&=&-\frac{\langle
g\bar{s}\sigma\cdot G
s\rangle}{3\cdot2^{6}\pi^{4}}\int_{\Lambda}^{1}
\frac{d\alpha}{\alpha}[3(1-\alpha)m_{c}+\alpha m_{s}]
(\alpha s-m_{c}^{2}),\\
\rho_{2}^{\langle\bar{q}q\rangle^{2}}&=&\frac{\langle\bar{q}q\rangle^{2}}{3\cdot2^{2}\pi^{2}}\int_{\Lambda}^{1}d\alpha[-(\alpha s-m_{c}^{2})+m_{s}m_{c}],\\
\rho_{2}^{\langle g^{3}G^{3}\rangle}&=&-\frac{\langle
g^{3}G^{3}\rangle}{3^{2}\cdot2^{12}\pi^{6}}\int_{\Lambda}^{1}d\alpha\bigg(\frac{1-\alpha}{\alpha}\bigg)^{3}(\alpha s-3m_{c}^{2}-6m_{s}m_{c}),\\
\rho_{2}^{\langle\bar{s}s\rangle\langle
g^{2}G^{2}\rangle}&=&\frac{m_{c}\langle\bar{s}s\rangle\langle
g^{2}G^{2}\rangle}{3^{2}\cdot2^{8}\pi^{4}}\int_{\Lambda}^{1}d\alpha\bigg[1+3\bigg(\frac{1-\alpha}{\alpha}\bigg)^{2}\bigg],\\
\hat{B}\Pi_{2}^{\mbox{cond}}&=&-\frac{m_{s}m_{Q}\langle\bar{q}q\rangle\langle g\bar{q}\sigma\cdot G q\rangle}{3\cdot2^{3}\pi^{2}}e^{-m_{Q}^{2}/M^{2}}\nonumber\\
&-&\frac{\langle\bar{q}q\rangle^{2}\langle\bar{s}s\rangle}{2\cdot3^{2}}\bigg(2m_{Q}+m_{s}+\frac{m_{s}m_{Q}^{2}}{M^{2}}\bigg)e^{-m_{Q}^{2}/M^{2}}\nonumber\\
&+&\frac{\langle g\bar{q}\sigma\cdot G q\rangle^{2}}{3\cdot2^{6}\pi^{2}}\bigg[-1-\frac{m_{Q}^{2}}{M^{2}}+\frac{m_{s}m_{Q}^{3}}{(M^{2})^{2}}\bigg]e^{-m_{Q}^{2}/M^{2}},
\end{eqnarray}
for the pseudoscalar diquark-pseudoscalar antidiquark case,
\begin{eqnarray}
\rho_{3}^{\mbox{pert}}&=&\frac{1}{3\cdot2^{8}\pi^{6}}\int_{\Lambda}^{1}d\alpha\bigg(\frac{1-\alpha}{\alpha}\bigg)^{3}(\alpha s-m_{c}^{2}+2m_{s}m_{c})(\alpha s-m_{c}^{2})^{3},\\
\rho_{3}^{\langle\bar{s}s\rangle}&=&-\frac{\langle\bar{s}s\rangle}{2^{5}\pi^{4}}\int_{\Lambda}^{1}d\alpha\frac{1-\alpha}{\alpha^{2}}[(1-\alpha)m_{c}-2\alpha m_{s}](\alpha s-m_{c}^{2})^{2},\\
\rho_{3}^{\langle g^{2}G^{2}\rangle}&=&-\frac{m_{c}\langle
g^{2}G^{2}\rangle}{3^{2}\cdot2^{9}\pi^{6}}\int_{\Lambda}^{1}d\alpha\bigg(\frac{1-\alpha}{\alpha}\bigg)^{3}[(2m_{c}-3m_{s})(\alpha s-m_{c}^{2})+m_{s}m_{c}^{2}],\\
\rho_{3}^{\langle g\bar{s}\sigma\cdot G s\rangle}&=&\frac{\langle
g\bar{s}\sigma\cdot G
s\rangle}{3\cdot2^{5}\pi^{4}}\int_{\Lambda}^{1}
\frac{d\alpha}{\alpha}[3(1-\alpha)m_{c}-2\alpha m_{s}]
(\alpha s-m_{c}^{2}),\\
\rho_{3}^{\langle\bar{q}q\rangle^{2}}&=&\frac{\langle\bar{q}q\rangle^{2}}{3\cdot2\pi^{2}}\int_{\Lambda}^{1}d\alpha[(\alpha s-m_{c}^{2})+2m_{s}m_{c}],\\
\rho_{3}^{\langle g^{3}G^{3}\rangle}&=&-\frac{\langle
g^{3}G^{3}\rangle}{3^{2}\cdot2^{10}\pi^{6}}\int_{\Lambda}^{1}d\alpha\bigg(\frac{1-\alpha}{\alpha}\bigg)^{3}(\alpha s-3m_{c}^{2}+3m_{s}m_{c}),\\
\rho_{3}^{\langle\bar{s}s\rangle\langle
g^{2}G^{2}\rangle}&=&-\frac{m_{c}\langle\bar{s}s\rangle\langle
g^{2}G^{2}\rangle}{3^{2}\cdot2^{7}\pi^{4}}\int_{\Lambda}^{1}d\alpha\bigg[1+3\bigg(\frac{1-\alpha}{\alpha}\bigg)^{2}\bigg],\\
\hat{B}\Pi_{3}^{\mbox{cond}}&=&-\frac{m_{s}m_{Q}\langle\bar{q}q\rangle\langle g\bar{q}\sigma\cdot G q\rangle}{3\cdot2\pi^{2}}e^{-m_{Q}^{2}/M^{2}}\nonumber\\
&-&\frac{\langle\bar{q}q\rangle^{2}\langle\bar{s}s\rangle}{3^{2}}\bigg(4m_{Q}-m_{s}-\frac{m_{s}m_{Q}^{2}}{M^{2}}\bigg)e^{-m_{Q}^{2}/M^{2}}\nonumber\\
&+&\frac{\langle g\bar{q}\sigma\cdot G q\rangle^{2}}{3\cdot2^{5}\pi^{2}}\bigg[1+\frac{m_{Q}^{2}}{M^{2}}+2\frac{m_{s}m_{Q}^{3}}{(M^{2})^{2}}\bigg]e^{-m_{Q}^{2}/M^{2}},
\end{eqnarray}
for the axial vector diquark-axial vector antidiquark case, and
\begin{eqnarray}
\rho_{4}^{\mbox{pert}}&=&\frac{1}{3\cdot2^{8}\pi^{6}}\int_{\Lambda}^{1}d\alpha\bigg(\frac{1-\alpha}{\alpha}\bigg)^{3}(\alpha s-m_{c}^{2}-2m_{s}m_{c})(\alpha s-m_{c}^{2})^{3},\\
\rho_{4}^{\langle\bar{s}s\rangle}&=&\frac{\langle\bar{s}s\rangle}{2^{5}\pi^{4}}\int_{\Lambda}^{1}d\alpha\frac{1-\alpha}{\alpha^{2}}[(1-\alpha)m_{c}+2\alpha m_{s}](\alpha s-m_{c}^{2})^{2},\\
\rho_{4}^{\langle g^{2}G^{2}\rangle}&=&-\frac{m_{c}\langle
g^{2}G^{2}\rangle}{3^{2}\cdot2^{9}\pi^{6}}\int_{\Lambda}^{1}d\alpha\bigg(\frac{1-\alpha}{\alpha}\bigg)^{3}[(2m_{c}+3m_{s})(\alpha s-m_{c}^{2})-m_{s}m_{c}^{2}],\\
\rho_{4}^{\langle g\bar{s}\sigma\cdot G s\rangle}&=&-\frac{\langle
g\bar{s}\sigma\cdot G
s\rangle}{3\cdot2^{5}\pi^{4}}\int_{\Lambda}^{1}
\frac{d\alpha}{\alpha}[3(1-\alpha)m_{c}+2\alpha m_{s}]
(\alpha s-m_{c}^{2}),\\
\rho_{4}^{\langle\bar{q}q\rangle^{2}}&=&\frac{\langle\bar{q}q\rangle^{2}}{3\cdot2\pi^{2}}\int_{\Lambda}^{1}d\alpha[-(\alpha s-m_{c}^{2})+2m_{s}m_{c}],\\
\rho_{4}^{\langle g^{3}G^{3}\rangle}&=&-\frac{\langle
g^{3}G^{3}\rangle}{3^{2}\cdot2^{10}\pi^{6}}\int_{\Lambda}^{1}d\alpha\bigg(\frac{1-\alpha}{\alpha}\bigg)^{3}(\alpha s-3m_{c}^{2}-3m_{s}m_{c}),\\
\rho_{4}^{\langle\bar{s}s\rangle\langle
g^{2}G^{2}\rangle}&=&\frac{m_{c}\langle\bar{s}s\rangle\langle
g^{2}G^{2}\rangle}{3^{2}\cdot2^{7}\pi^{4}}\int_{\Lambda}^{1}d\alpha\bigg[1+3\bigg(\frac{1-\alpha}{\alpha}\bigg)^{2}\bigg],\\
\hat{B}\Pi_{4}^{\mbox{cond}}&=&-\frac{m_{s}m_{Q}\langle\bar{q}q\rangle\langle g\bar{q}\sigma\cdot G q\rangle}{3\cdot2\pi^{2}}e^{-m_{Q}^{2}/M^{2}}\nonumber\\
&-&\frac{\langle\bar{q}q\rangle^{2}\langle\bar{s}s\rangle}{3^{2}}\bigg(4m_{Q}+m_{s}+\frac{m_{s}m_{Q}^{2}}{M^{2}}\bigg)e^{-m_{Q}^{2}/M^{2}}\nonumber\\
&+&\frac{\langle g\bar{q}\sigma\cdot G q\rangle^{2}}{3\cdot2^{5}\pi^{2}}\bigg[-1-\frac{m_{Q}^{2}}{M^{2}}+2\frac{m_{s}m_{Q}^{3}}{(M^{2})^{2}}\bigg]e^{-m_{Q}^{2}/M^{2}},
\end{eqnarray}
for the vector diquark-vector antidiquark case. For brevity,
the light quark condensate $\langle\bar{u}u\rangle$ or $\langle\bar{d}d\rangle$
is uniformly represented by $\langle\bar{q}q\rangle$,
and the integration limit is defined as
$\Lambda=m_{c}^2/s$.
Some high dimension condensates are included above,
e.g. $\langle\bar{q}q\rangle\langle g\bar{q}\sigma\cdot G q\rangle$,
$\langle\bar{q}q\rangle^{2}\langle\bar{s}s\rangle$,
and $\langle g\bar{q}\sigma\cdot G q\rangle^{2}$.
In the chosen work windows, the relative contributions of these kind condensates are
very small, and they merely affect the final results in some sort.
Most of other high dimension condensates are formed
into some similar terms with negative powers of $q^{2}$, which will be strongly suppressed
after making a Borel transform.
Similarly, from the work \cite{tetra5} on studying $D_{s0}^{*}(2317)$
as a $qs\bar{q}\bar{c}$ tetraquark,
one can see that high dimension
contributions are rather small.


\end{document}